\documentclass{article}

\usepackage{chicago}
\usepackage{amssymb}
\usepackage{amsmath}
\usepackage{graphicx}

\newcommand{\iec}{\mbox{i.\,e.\,}}

\newcommand{\vctr}[1]{\ensuremath{\mathbf{ #1 }}}

\newcommand{\dr}[1]{\ensuremath{\mathrm{d} #1\,}}

\newcommand{\dbd}[2]{\ensuremath{\frac{\dr{#1}}{\dr{#2}}}}

\newcommand{\pbp}[2]{\ensuremath{\frac{\partial #1}{\partial #2}}}

\newcommand{\op}[1]{\ensuremath{\widehat{\textsf{\ensuremath{#1}}}}}

\newcommand{\be}{\begin{equation}}
\newcommand{\ee}{\end{equation}}

\renewcommand{\vctr}[1]{\ensuremath{\overrightarrow{\mathbf{ #1 }}}}

\def\stkd{{\mathchar'26\mkern-12mu d}}
\begin{document}

\title{On relativistic thermodynamics}
\author{David Wallace\thanks{Department of History and Philosophy of Science / Department of Philosophy, University of Pittsburgh; \texttt{david.wallace@pitt.edu}}}
\maketitle

\begin{abstract}
‘Relativistic thermodynamics’ should be understood not as a generalization of a non-relativistic theory but as an application of a general thermodynamic framework, neutral as to spacetime setting and allowing arbitrary conserved quantities, to the specific case of relativity. That framework gives an unambiguous result as to the thermodynamics of relativistically moving systems (an answer coinciding with Einstein’s, and Planck’s, original results.) Thermodynamic temperature is unambiguously defined as rate of change of energy with entropy at constant momentum; that said, its operational significance is limited and other measures of energy/entropy covariance, which incorporate momentum transfer, are often more useful.
\end{abstract}

\section{Introduction}

Relativistic thermodynamics is almost as old as relativity itself and yet remains surprisingly controversial. Liu's \citeyear{liurelativisticthermodynamics} history of the subject concludes by describing the theory as `one of the most recalcitrant in resisting the efforts of relativization'; in recent work \citeN{chuaTfallsapart} goes further in claiming that  relativistic thermodynamics leads to `a breakdown of the classical non-relativistic concept of temperature'. The issue has acquired a new urgency in the context of recent philosophical criticism of the longstanding claims of analogy between black hole behavior and thermodynamics\footnote{Notably by Chua (ibid.) in passing, and by \citeN{doughertycallender} --- though see also the response to the latter in Wallace~\citeyear{wallaceblackholethermodynamics}.}.

It is at first sight surprising that any such controversies are compatible with the state of modern thermal physics. There is nothing obviously non-relativistic about modern thermodynamics, or the statistical mechanics that underpins it: to the contrary, it is absolutely routine to apply thermodynamics to systems --- the plasma in a fusion reactor, the interior of the Sun, the shock front of a supernova, Big Bang nucleosynthesis --- which are not even faintly `non-relativistic'. At an elementary level, thousands of physics undergraduates calculate the thermodynamic properties of black-body radiation every year without any suggestion that doing so involves a relativistic ingredient not present in similar calculations for the ideal gas; at a more advanced level, one will search the index of monographs on finite-temperature quantum field theory or relativistic astrophysics in vain for any such suggestion.\footnote{I searched the indexes of \citeN{battaner}, \citeN{kapustagale}, and \citeN{padmanabhanastrophysics}.}

Things become clearer upon noting the main locus of the controversy: the thermodynamics of systems in relative motion, and the transformation properties of temperature and other thermodynamic quantities under Lorentz transformations, which indeed play little role\footnote{Not no role: observational astrophysics often requires us to consider what a system at thermal equilibrium in one frame looks like in another, a point to which I return in section \ref{covariant-thermodynamics}.} in the application of thermodynamics to relativistic systems (where thermal calculations are almost always carried out in a local rest frame). Here one sees the possibility of a controversy that could persist without troubling the physics mainstream. And controversy there is: a proposed relativistic thermodynamics due to Planck and Einstein was initially agreed upon by the physics community but that agreement collapsed in the 1950s, since when a voluminous literature has developed but no consensus has been restored. 

A consistent theme of this literature is that the project of `relativistic thermodynamics' is the project of starting with the thermodynamics of systems at rest and working out how to  generalize it to relativistically moving systems. Common strategies include (i) beginning with the classic statement of the First Law for a fluid,
\be\label{classic-first-law}
\mathrm{d}U=\stkd Q+\stkd W \equiv T\mathrm{d}S - P \mathrm{d}V,
\ee
and seeking appropriate relativistic transformation laws for its various terms, and (ii) going back to the operational foundations of thermodynamics and seeking a relativistic generalization of the Carnot cycle, appropriate for running heat engines between relatively moving systems. Since both (i) and (ii) appear ambiguous, with multiple plausible-looking transformation laws and multiple intuitively-reasonable definitions of a Carnot cycle, unsurprisingly this has led to underdetermination as to the `correct' relativistic extension, and hence to the possibilities either that there are multiple intertranslatable extensions or that there is no fully satisfactory relativistic extension and the core concepts of thermodynamics break down in relativistic contexts.

The core argument of this paper is that no relativistic \emph{extension} of thermodynamics is required, because standard thermodynamics already has sufficient scope to handle moving systems. As I review in section~\ref{general-thermodynamics}, the form of thermodynamics expressed by ($\ref{classic-first-law}$) is a narrow special case of a more general framework, applicable only to systems where energy is the only conserved quantity and volume is the only externally-set parameter. Thermodynamics in general handles arbitrary choices of conserved quantities over and above energy, as well as a wide class of external parameters beyond volume. This more general formalism has been known for more than a century, has been applied extremely widely, and is uncontroversial.\footnote{More precisely: it inherits the many controversies of thermal physics but adds no new ones.} It gives an unambiguous definition of thermodynamic temperature: it is the rate of change of entropy with energy while all other conserved quantities and external parameters are held constant.

Obtaining the thermodynamics of moving systems just requires us to observe that they fall into this more general framework, since they conserve \emph{momentum} as well as energy, and then to apply that framework. No new conceptual insights are required: we just need to turn the handle of the thermodynamic formalism. I carry this out in section \ref{thermodynamics-of-moving-bodies} (both for Poincar\'{e}-covariant and Galilean-covariant systems). But the relativistic covariance of the resultant theory is somewhat obscured, essentially because the idea of exchanging energy at fixed momentum is not an invariant concept and has limited operational significance (as opposed to, say, exchanging energy at fixed particle number or charge); I develop this point, and present a more covariant version of the theory, in section~\ref{covariant-thermodynamics}. In that section I define various `generalized temperatures' that measure how energy varies with entropy under assumptions other than the constancy of momentum, such as `constant-velocity temperature' (which measures the rate of change of energy with entropy at constant velocity) and `radiation temperature', which measures how energy covaries with entropy when it is emitted as radiation. These quantities are physically useful in various contexts; nonetheless, our ability to define them is just a matter of convenience and does not imply any indeterminacy in the formulation of relativistic thermodynamics. I develop this point in some detail in section~\ref{constant-velocity-thermodynamics} (where I argue that we cannot take the velocity of a moving system to be an external control parameter like the volume of a box, and so `constant-velocity temperature' cannot be understood as a valid form of thermodynamic temperature) and section~\ref{rest-temperature} (where I consider the relation between thermodynamic temperature and `rest temperature', the temperature of a system in its rest frame, and analogize it to the relation between inertial mass and rest mass). In section~\ref{stat-mech} I consider the statistical-mechanical underpinnings of the thermodynamics of moving systems and argue that just as with thermodynamics, the extant framework of equilibrium statistical mechanics is already broad enough to include moving systems and to give unambiguous predictions as to their statistical-mechanical representation.

Most of the detailed formulae in the paper can be found in various bits of the relativistic-thermodynamics literature, although the method by which I derive and interpret them has not (so far as I am aware) been previously discussed. For the sake of logical clarity, the main part of the paper is self-contained and I do not attempt to relate specific results to the extant literature. In the final section, however (section \ref{what-has-gone-before}), I review the historical debate and place my results in historical context, observing specifically that they essentially vindicate the original Planck-Einstein proposals, although their methods for deriving them are quite different from mine. In this concluding section I reprise, with the benefit of the results derived in the main part of the paper, the contrast I described above: between the approach taken in the historical literature, which takes relativistic thermodynamics as a novel extension of an existing theory of stationary systems, and the approach of this paper, which takes it as simply an application of well understood concepts.

\subsection*{Notation}

I use units where $c=k_B=1$. 

$\vctr{x}$ denotes a 3-vector; $\tilde x$ denotes a 4-vector; $(U,\vctr{p})$ denotes a 4-vector expressed relative to some inertial frame in which its 0 component is $U$ and its spatial component is $\vctr{p}$. I assume a timelike-negative metric, use Greek subscripts and superscripts to denote the indices of 4-vectors, and assume the Einstein summation convention for those indices. I fix a specific frame which I call the `lab' frame; unless otherwise stated, relativistically non-invariant expressions should be understood relative to this frame.   (So, for instance, if I say without qualification that a body is `moving' or `at rest', these are to be understood relative to the lab frame.) 

In general I use the symbols $U,\vctr{p},\vctr{v},M$ to refer respectively to the energy, momentum, velocity and rest mass of a body. These are of course interrelated: standard relativistic kinematics tell us, for instance, that $\vctr{p}=\vctr{v}U$ and that $M^2=U^2 - \vctr{p}\cdot \vctr{p}$. In many cases I will take a subset of these variables (usually $U$ and $\vctr{p}$, occasionally $M$ and $\vctr{v}$) as independently specified and regard the others as functions of them; to avoid cluttering the notation I do not make this dependence explicit. (So if I say, for instance, that a body has 4-momentum  $(U,\vctr{p})$ and then refer to its velocity $\vctr{v}$, I am suppressing a functional dependence $\vctr{v}=\vctr{v}(U,\vctr{p})=\vctr{p}/U$.)
I write $\tilde v$ for the  4-velocity and  $\tilde p$ for the  4-momentum. 

The function $v\rightarrow\gamma(v)$ is as usual defined as $\gamma(v)=(1-v^2)^{-1/2}$. Again to avoid cluttering the notation, where we are considering a body with velocity $\vctr{v}$ I suppress the functional dependence of $\gamma$ on $|\vctr{v}|$: by definition $\gamma \equiv \gamma(|\vctr{v}|)$.

\section{General thermodynamics}\label{general-thermodynamics}

The foundation of equilibrium thermodynamics (called the ‘minus first law’ by \citeN{BrownUffink2001}) is that isolated systems evolve towards unique equilibrium states. 

But what does ‘unique’ mean here? If ‘isolated’ means that energy does not flow into or out of a system during equilibration, then of course different-energy systems will obtain different equilibrium states. But if in addition there are other conserved quantities than energy, and if ‘isolated’ means that these too cannot be exchanged with the environment, then equilibrium states will be individuated by the values of those other conserved quantities as well as by energy. And if the system’s dynamics depends on some externally controlled variable --- like the volume, for instance --- and if that variable is held fixed during equilibration, then different values of that variable lead to different equilibria.\footnote{This latter point is recognized in Brown and Uffink’s precise statement of the minus first law: ``An isolated system in an arbitrary initial state \emph{within a finite fixed volume} will spontaneously attain a unique state of equilibrium’’ (my emphasis). But volume is not the only possible external parameter, and even for fixed volume there may be conserved quantities other than energy.}

Examples are widespread. In chemical thermodynamics there are conservation laws tracking the separate conservation of each element; in nuclear chemistry element number is not conserved but quantities like charge and baryon number are; in the thermodynamics of magnetic matter volume is normally fixed and the role of external parameter is played by a magnetic field. The thermodynamics of a hot rock involves no external parameters and no conserved quantities except energy; the thermodynamics of a box of hydrogen atoms involves energy, volume, and number of atoms; the thermodynamics of a box of black-body radiation involves only energy and volume, since photon number is not conserved.

The formalism of thermodynamics is wide enough to incorporate all these cases and more. Let us denote any conserved quantities, other than the energy $U$, by $N_1,\ldots N_K$, and any external parameters by $V_1,\ldots V_M$. The second law of thermodynamics then amounts to the statement that there is a (piecewise smooth) function \[S(U,N_1,\ldots, N_K,V_1,\ldots, V_M)\] --- the \emph{thermodynamic entropy} --- of the conserved quantities and external parameters such that (1) if an isolated system initially at equilibrium is allowed to evolve under externally-induced time dependence of its external parameters and then to return to equilibrium, the value of $S$ will not have decreased; and (2) if two or more systems initially at equilibrium are dynamically coupled so as to be able to exchange energy and other additive conserved quantities, and then the coupling is removed and they are allowed to come to equilibrium, then again the total value of $S$ will not have decreased. (It is common in foundational work to present the second law in more directly operational terms, but in practical applications what matters is the entropy form I state here.)

Differentiating, we obtain
\be\label{entropy1}
\mathrm{d}S = \beta \mathrm{d}U + \sum_{i=1}^K \theta_i \mathrm{d}N_i + \sum_{i=1}^M \alpha_i \mathrm{d}V_i
\ee
where $\beta$, $\theta_i$ and $\alpha_i$ are all functions of $U$, the $N_i$, and the $V_i$, given explicitly by
\be
\beta=\left(\pbp{S}{U}\right)_{N_j,V_j}\,\,\,\,\theta_i=\left(\pbp{S}{N_i}\right)_{U,N_j(i\neq j),V_j}\,\,\,
\alpha^i=\left(\pbp{S}{V_i}\right)_{U,N_j,V_j(i\neq j)}.
\ee
As stated this is an entirely formal statement about the space of equilibrium states, but it has an operational interpretation if we take $\mathrm{d}U$, $\mathrm{d}N_i$, and $\mathrm{d}V_i$ to be small but finite changes to the constants and parameters caused by some external intervention (with the system otherwise being kept isolated). In the limit of small changes, $\mathrm{d}S$ becomes the entropy change caused by that intervention once the system returns to equilibrium, and the second law becomes the requirement that that change is nonnegative so long as the system is isolated from its environment. The various parameters $\beta$, $\theta_i$, $\alpha_i$ then parameterize how entropy covaries with each of the conserved quantities and parameters as the others are held constant. In particular, the \emph{inverse temperature} $\beta$ is the rate of change of entropy with energy under infinitesimal changes that leave constant the other conserved quantities and all the parameters.

Partly for historical reasons, it is standard to rewrite (\ref{entropy1}) as
\begin{eqnarray}\label{firstlaw}
\mathrm{d}U &=& \frac{1}{\beta}\mathrm{d}S + \sum_i\left(-\frac{\theta_i}{\beta}\right)\mathrm{d}N_i - \sum_i\left(\frac{\alpha_i}{\beta}\right)\mathrm{d}V_i\nonumber \\
&\equiv&T \mathrm{d}S + \sum_i \mu_i \mathrm{d}N_i - \sum_i P_i \mathrm{d}V_i.
\end{eqnarray}
This expression is sometimes called the \emph{First Law}, and I follow this convention here (without prejudice as to what connection it bears to the historical First Law). $T=1/\beta$ is the \emph{thermodynamic temperature}; $\mu_i=-\theta_i/\beta$ is a \emph{generalized potential}; $P_i=\alpha_i/\beta$ is a \emph{generalized pressure}. The operational significance of $T$ in isolation can be understood by considering processes in which energy, but no other conserved quantity, is allowed to flow between two systems (while holding their parameters fixed); in this case we can swiftly read off the standard thermodynamic principle that spontaneous flow is possible only from a higher-temperature to a lower-temperature system, and (with a little more algebra) that the efficiency of a heat engine that works by transferring energy \emph{but nothing else} between two systems at temperatures $T_1,T_2$ is bounded by the Carnot efficiency $(1-T_2/T_1)$. The generalized potentials have somewhat analogous operational meanings: for instance, we can transfer particle number reversibly between two systems at the same temperature and extract energy in doing so iff their chemical potentials differ.

For a philosophically sensitive review of these ideas, see \cite{wallaceirreversibility}; for a straightforward review of the physics, see \cite{callen} or any other good graduate-level textbook on thermodynamics.

\section{The thermodynamics of moving bodies}\label{thermodynamics-of-moving-bodies}

The canonical examples of thermodynamic systems conserve energy but not momentum, and the reason is simple: these canonical systems are confined in some kind of external box, and their constituents literally bounce off the walls, transferring momentum to them. No system of this kind can have dynamics that is covariant under velocity boosts: put simply, if the fluid in a stationary box is moving, it will slam into the walls of the box, losing its momentum in the process (and will then equilibrate at zero momentum). A thermodynamic system that can be in motion must include its box, if any, as part of the system itself; such a system will conserve momentum as well as energy, and so the momentum must be included along with energy on the list of conserved quantities characterizing the system.

We can also see the need to include momentum as well as energy in a covariant system directly through considerations of its covariance. Energy is not a (Galilean or special-relativistic) scalar: under boosts, it mixes with momentum, and so `this system conserves energy but not momentum' is a frame-dependent notion (the frame in question normally being the rest frame of the box confining the system). A covariant system conserves both, and so both must be included in the characterization of the space of equilibrium states. 

(A complication arises. Equilibrium is traditionally described as the state a system reaches when all of its macroscopic degrees of freedom are unchanging. But of course a moving body is, well, moving, and precisely because momentum is conserved, that movement does not cease at equilibrium; similarly, in general a rotating body will tumble in space, and that tumbling will not cease as long as angular momentum is conserved.

Nonetheless there clearly is a physically relevant sense of equilibration here: in a cylinder of gas in empty space, tumble and fly though it might, the contents will still reach an appropriately steady state. We can characterize that sense more precisely: suppose (assuming for definiteness classical Lagrangian mechanics) that the system has coordinates $q^1,\ldots q^N$ but that the Lagrangian does not depend on $q^1$, so that translation in $q^1$ is a symmetry. (At least locally, any configuration symmetry can be so expressed.) Then the conjugate momentum 
\be
p_1=\pbp{L}{q^1}
\ee
is conserved, and there is a self-contained dynamics for the remaining coordinates $q^2,\ldots q^N$, in which $p_1$ can be treated as a time-independent parameter in the expression for the Hamiltonian. Equilibrium can now be understood with respect to these coordinates. For instance, for a translation-invariant $N$-particle system we can take the $N-3$ translationally invariant degrees of freedom to collectively reach equilibrium.)

Following the general framework discussed in section \ref{general-thermodynamics}, let's consider a system which conserves both energy and momentum, and which in addition has one externally controlled parameter $V_r$, with conjugate generalized pressure $\lambda$. (I have in mind that $V_r$ is the system volume in its rest frame, but little hangs on this; generalization to two or more, or no, external parameters is straightforward.) The general form of the First Law for such a system is
\be
\mathrm{d}U = T \mathrm{d}S - \lambda \mathrm{d}V_r + \vctr{\mu}\cdot\mathrm{d}\vctr{p}
\ee
with $\vctr{\mu}$ being a vector of potentials conjugate to the coordinates of momentum.

In the thermodynamics of fluids, thermodynamic pressure is by definition (minus) the rate of change of energy with volume at constant entropy. But of course it can be identified with \emph{mechanical} pressure (that is: the force per unit area normal to the comoving walls of the fluid's container), through the obvious and familiar argument that if a small section of a fluid's wall with area $\delta A$ is moved away from the fluid a distance $\delta x$, then the mechanical work done by the fluid is 
\[\mbox{(mechanical pressure) }\times \delta A \times \delta x.\]
Similarly, the thermodynamic potential conjugate to momentum is by definition the rate of change of system energy with momentum at constant entropy, but it also has a mechanical interpretation. To see this, suppose that a small force $\vctr{f}$ acts on the system over some time $\delta t$, hence changing the system's momentum by $\vctr{f}\delta t$. This will move the system out of equilibrium (for instance, if the system is a box of fluid then if we push on the box then its moving walls will agitate the fluid within) but it will quickly reequilibrate, and in the limit as the rate at which the system is pushed becomes arbitrarily small, the system will remain arbitrarily close to equilibrium and the increase in $S$ will be arbitrarily low. In the limit, the First Law gives us
\be
\delta U = \vctr{\mu}\cdot \vctr{f}\delta t.
\ee
But it is also true that if the system has velocity $\vctr{v}$, then the force is applied over a distance $\vctr{v}\delta t$, so that the mechanical work done on the system is
\be
\stkd W = \vctr{v} \delta t \cdot \vctr{f}
\ee
and equating these two tells us that the potential conjugate to momentum is just velocity, so that the First Law can be rewritten as
\be\label{first-law-rest-frame}
\mathrm{d}U = T \mathrm{d}S - \lambda \mathrm{d}V_r + \vctr{v}\cdot\mathrm{d}\vctr{p}.
\ee

Since this transformation on the system is just the active implementation of a velocity boost, it also follows that entropy $S$ transforms as a scalar, and we can use this fact, combined with some relativistic kinematics, to derive the transformation rules for other thermodynamic quantities. Specifically, if entropy is a scalar it can depend on $U$ and $\vctr{p}$ only through $M$, $S=S(M,V_r)$. At rest, $M=U$ and $\vctr{v}=0$, so that the First Law reduces to
\be
\mathrm{d}M = T_r \mathrm{d}S - P_r \mathrm{d}V_r
\ee
where I write $T_r$ for the \emph{rest temperature}, the temperature of the system in its own rest frame, and likewise $P_r$ for the \emph{rest pressure}. This expression relates scalar quantities and so is itself valid in any inertial frame. Next, note that
\be
M\mathrm{d}M = U \mathrm{d}U - \vctr{p}\cdot \mathrm{d}\vctr{p}
\ee
and so
\be\label{mass-diff}
\mathrm{d}M = \gamma \mathrm{d}U - \gamma\vctr{v}\cdot \mathrm{d}\vctr{p}.
\ee
Equating the expressions (\ref{first-law-rest-frame}) and (\ref{mass-diff}) and rearranging gives
\be\label{antithesis-firstlaw-2}
\mathrm{d}U = \frac{T_r}{\gamma}\mathrm{d}S - \frac{P_r}{\gamma}\mathrm{d}V_r + \vctr{v}\cdot \mathrm{d}\vctr{p}.
\ee
Comparing this with the equation of state (\ref{first-law-rest-frame}) tells us that:
\begin{itemize}
\item The temperature of a system moving with velocity $\vctr{v}$ is $1/\gamma$ times its rest temperature.
\item The potential conjugate to rest volume is $1/\gamma$ times the rest pressure.
 \end{itemize}

 The operational significance of temperature is exactly what it is in standard thermodynamics (because, to repeat: this all just is standard thermodynamics). Specifically, it determines the efficiency of a heat engine working between two systems that transfers energy but no other conserved quantity. Such a heat engine, operating between systems at velocities $\vctr{v}_1$, $\vctr{v}_2$ and with rest temperatures $T_{r1},T_{r2}$, has efficiency
 \be
 e\leq 1 - \frac{T_{1r}/\gamma(|\vctr{v}_1|)}{T_{2r}/\gamma(|\vctr{v}_2|)}.
 \ee
 The fact that temperature is rate of change of energy with entropy \emph{at constant momentum} is crucial to resolving an apparent paradox, akin to the classic ``paradoxes'' of relativistic kinematics: how can it be that for two systems in relative motion and with the same rest temperature, an observer comoving with each system will agree that the other system is colder? To resolve this paradox, suppose that system $A$ is at rest in the lab frame, and that it transfers a quantity of energy $\delta U$ (but no momentum) to another system, $B$, moving at velocity $\vctr{v}$. If both systems have rest temperature $T_r$, this is entropically favorable: system $A$'s entropy decreases by $\delta U/T$ and system $B$'s increases by $\gamma \delta U/T$.

 In the frame of system $B$, it would be entropically unfavorable just to transfer some energy from system $A$, since $A$ is at a lower temperature. But the original transfer is \emph{not} just a transfer of energy, but also of  momentum: in fact, the original transfer described in $B$'s frame is  a transfer of energy $\gamma \delta U$ and of momentum $-\gamma \vctr{v}\delta U$. Rearranging (\ref{antithesis-firstlaw-2}) (and setting $\mathrm{d}P_r=0$) gives
 \be
 \mathrm{d}S = \frac{\gamma}{T}\left( \mathrm{d}U + \vctr{v}\cdot \mathrm{d}\vctr{p}\right).
 \ee
 The decrease in entropy of system $A$ (which has velocity $-\vctr{v}$ in $B$'s frame) is then $\delta U/T$, while the increase in entropy of system $B$ (which has velocity $0$ in its own frame) is $\gamma \delta U/T$; in both cases reproducing the results already obtained in $A$'s frame. 
 
 As for the operational significance of the $(P_r/\gamma)\mathrm{d}V_r$ term, expanding the box at constant momentum will change its size through two mechanisms: (i) its rest volume obviously increases, (ii) extracting energy from a system at constant momentum increases its speed, and hence the level of Lorentz contraction. But if we assume that the machinery of the box has large rest mass compared to the fluid in the box then the latter factor can be neglected. Since $\delta U=-(P_r/\gamma) \delta V_r$ and $V_r = \gamma V$, in this regime we have $\delta U = - P_r \delta V$. Since mechanical pressure is invariant under Lorentz boosts\footnote{There are many ways to see this; one simple and fairly physical way is to consider the force required to confine a fluid of rest-frame pressure $P$ inside a cubical box of rest-frame face area $A$, which at rest is $PA$, and then consider how force and area change under Lorentz boosts. If the box is moving along the $x$ axis, the 4-force required on a surface normal to that axis is $(0,PA,0,0)$ in the box rest frame, and so $(\gamma  v PA, \gamma PA, 0,0)$ in the lab frame; since force is rate of change of 4-momentum with proper time and force is rate of  change of momentum with coordinate time, the force is invariant and hence so is the pressure. The 4-force required on a surface parallel to the $x$ axis (say, for definiteness, in the $xz$ plane) is $(0,0,PA,0)$ in the box rest frame and so unchanged in the lab frame; the force is thus reduced by a factor $\gamma$ but so (via Lorentz contraction) is the area of the surface, so that the pressure is again unchanged.} this just replicates the equality of thermodynamic and mechanical pressure.
 
  Incidentally, if we assume instead \emph{Galilean} physics, then $S$ becomes a function of $V_r$ and of the center-of-mass energy $U_{CM}$, with
 \be
 U=U_{CM}+\frac{\vctr{p}\cdot \vctr{p}}{2M}
 \ee
 (where of course the mass $M$ is now independent of the center-of-mass energy) and hence
 \be
 \mathrm{d}U=\mathrm{d}U_{CM}+\vctr{v}\cdot \mathrm{d}\vctr{p}
 \ee
  from which we derive
\be\label{galilean-firstlaw}
\mathrm{d}U = T_r\mathrm{d}S - P_r\mathrm{d}V_r + \vctr{v}\cdot \mathrm{d}\vctr{p}
\ee  
--- that is, temperature and pressure are invariant under boosts in the nonrelativistic limit (something we could also have obtained from the low-volume limit of (\ref{antithesis-firstlaw-2})).
 
 \section{Covariant thermodynamics and generalized temperature}\label{covariant-thermodynamics}
 
  The relativistic covariance of this form of thermodynamics is somewhat obscured in (\ref{antithesis-firstlaw-2}) --- unsurprisingly, since it treats energy and momentum differently. Relatedly, the operational significance of temperature may be well-defined, but it is a less natural quantity than its analogs in other forms of thermodynamics. In, say, a fluid of variable mass, it is extremely natural to consider processes in which heat but not particles can be exchanged between systems. But the notion of a process in which energy but not momentum is conserved is not relativistically invariant, and so its operational significance is lessened.
  
  To obtain a more manifestly covariant version of relativistic thermodynamics, start with $S=S(M,V_r)$, differentiate it  and use the covariant version of (\ref{mass-diff}),
  \be
  M \mathrm{d}M = - \tilde{p}_\mu \mathrm{d}\tilde{p}^\mu
  \ee
  to obtain
 \begin{eqnarray}
 \mathrm{d}S &=& -\tilde{v}_\mu\left(\pbp{S}{M}\right)_{V_r}\mathrm{d}\tilde{p}^\mu + \left(\pbp{S}{V_r}\right)_{M}\mathrm{d}V_r\nonumber \\
 &\equiv&-\tilde\beta_{\mu} \mathrm{d}\tilde{p}^\mu + P_r \mathrm{d}V_r
 \end{eqnarray}
 where $\tilde\beta$ is the \emph{inverse 4-temperature},
 \be
 \tilde\beta= \frac{1}{T_r}\tilde{v},
 \ee
 and is definitionally the rate of change of entropy with 4-momentum. 
 
This differential form of the equation of state is equivalent to the First Law (\ref{antithesis-firstlaw-2}) but makes its covariance manifest: we can now see that the thermodynamic temperature transforms as the inverse of the time component of a 4-vector.

To get further insight into this form of the First Law, suppose that a quantity $\delta U$ of energy is transferred to a moving system, and that the velocity of the transferred energy is $\vctr{u}$, \iec its momentum is $\vctr{u}\delta U$. The change in entropy (assuming no change in rest volume) is
\be
\delta S = \frac{\gamma}{T_r}(1 - \vctr{u}\cdot\vctr{v})\delta U.
\ee
We can define the \emph{generalized temperature} for velocity $\vctr{u}$ as
\be
T(\vctr{u})=\frac{T_r}{\gamma(1-\vctr{u}\cdot \vctr{v})}.
\ee
It is the rate of change of energy with entropy for velocity-$\vctr{u}$ energy transfers. And so it determines the maximum efficiency of a heat engine acting between systems that works by transferring energy at velocity $\vctr{u}$.

For example:
\begin{enumerate}
\item If we set $\vctr{u}=0$, the generalized temperature is the rate of change of energy with entropy at constant momentum, which of course is just the thermodynamic temperature.
\item If we set $\vctr{u}=\vctr{v}$, the generalized temperature is the \emph{constant-velocity temperature} $T_v$, the rate of change of energy with entropy at constant velocity (since if the energy transferred to or from a system has the same velocity as the system, that velocity will be left unchanged by the transfer). It is given by
\be
T_v = \gamma^2 T = \gamma T_r;
\ee
notice that unlike the thermodynamic temperature, it increases rather than decreases with system velocity.
\item If we set $|\vctr{u}|=1$, so that the energy transferred is in the form of radiation, the generalized temperature is the \emph{radiation temperature} $T_{rad}$. It is a function of the angle $\theta$ between $\vctr{u}$ and $\vctr{v}$: its specific form is
\be
T_{rad}(\theta) = \frac{T_r}{\gamma(1-v \cos \theta)}.
\ee
Not coincidentally, $\gamma(1-v \cos \theta)$ is the Doppler shift factor for radiation emitted from a moving body; the radiation temperature is the observed temperature of a relativistically moving source of thermal radiation, a fact well known in astrophysics.
\end{enumerate}

\section{Constant-velocity thermodynamics?}\label{constant-velocity-thermodynamics}

Entropy is a function of energy, momentum, and rest volume, and momentum is in turn a function of energy and velocity, so of course we can rearrange the First Law to express $\mathrm{d}U$ in terms of $\mathrm{d}\vctr{v}$ instead of $\mathrm{d}\vctr{p}$. Indeed, if we insert $\mathrm{d}\vctr{p}=U\mathrm{d}\vctr{v}+\vctr{v}\mathrm{d}U$ into (\ref{antithesis-firstlaw-2}) and rearrange, we get
\be\label{alt-first-law}
\mathrm{d}U=\gamma T_r \mathrm{d}S - \gamma P_r \mathrm{d}V_r +\gamma^2 \vctr{p}\cdot \mathrm{d}\vctr{p}.
\ee
Note that the coefficient in front of $\mathrm{d}S$ is the constant-velocity temperature, just as we would expect from its definition. So: what prevents us from interpreting relativistic temperature as constant-velocity temperature, and (\ref{alt-first-law}) as the true statement of the First Law?

There are actually two ways to read this proposal. The first is to hold on to the formulation of thermodynamics I gave in section \ref{thermodynamics-of-moving-bodies}, and simply to \emph{define} temperature as constant-velocity temperature. Certainly we can do this if we want. We can define `temperature' as the rate of change of energy with entropy at constant velocity if we want to. We can define `temperature' as any other of the generalized temperatures defined in section \ref{covariant-thermodynamics} if we want to. We can even define `temperature' as the average lifespan of the Amazonian marmoset if we want to. It's a free country. But this trivial fact about language is irrelevant to the fact that `temperature', as it is \emph{actually defined} in modern equilibrium thermodynamics, is rate of change of energy with entropy while all other conserved quantities and external parameters are held constant. There is nothing substantive about the proposal to use `temperature' to refer to constant-velocity temperature; it is just a proposal to redefine our terminology.

Nor is there anything particularly relativistic about this proposal. Consider some system with conserved quantities $N_i$ and associated potentials $\mu^i$. We can perfectly well ask how energy covaries with energy at constant potential rather than constant $N_i$, and we can define the \emph{constant-potential temperature} $T_\mu$ as the rate of change of energy at constant potential. Indeed, if we increase the entropy by $\delta S$ and then adjust the $N_i$ so as to keep the $\mu_i$ constant, we will have
\be
\delta N_i = \left(\pbp{N_i}{S}\right)_\mu \delta S
\ee
and so
\be
\delta U = \left( T + \sum_i \mu^i \left(\pbp{N_i}{S}\right)_\mu \right)\delta S
\ee
so that
\be\label{generalized-temperature}
T_\mu = T + \sum_i \mu^i \left(\pbp{N_i}{S}\right)_\mu.
\ee
So $T_\mu$ will in general differ from $T$, and (depending on the particular thermodynamic context) might be a physically important quantity, but it is not, definitionally, the thermodynamic temperature.

(Note that if we specialize back to relativistic thermodynamics, (\ref{generalized-temperature}) becomes
\be
T_v=T + \vctr{v}\cdot \left(\pbp{\vctr{p}}{S}\right)_{\vctr{v}} =\frac{T_r}{\gamma}+v^2\gamma \pbp{M}{S}= \frac{T_r}{\gamma}+v^2\gamma T_r=\gamma T_r,
\ee
reproducing our previous results for the constant-velocity temperature.)

There  is a more interesting way to try to read (\ref{alt-first-law}) as the First Law, and the constant-velocity temperature as the temperature. This is to suppose that we interpret velocity $\vctr{v}$ not as definitionally $\vctr{p}/U$ but as a set of external parameters, akin to the rest volume of the box. Equivalently, we might try to formulate relativistic thermodynamics not in terms of one parameter $V_r$ and four conserved quantities $\vctr{p},U$ but in terms of four parameters $V_r,\vctr{v}$ and one conserved quantity $U$.

At first sight, this can be made to work. Recall that we can model a fluid at rest as confined within a region by a potential barrier, so that the volume of the region becomes a parameter controlling the potential function: if the box occupies some region $R$, then the potential function is zero inside $R$ and climbs rapidly to a very high value outside $R$. 

For the system to be in motion, then, is for the potential itself to be in motion, confining the system to a time-dependent region. Note that the system does not conserve momentum, since collisions with the confining potential change the system momentum, so our treatment appears self-consistent: the only conserved quantity is energy. And of course the First Law for a system modelled this way indeed takes the form (\ref{alt-first-law}).

But there is a fatal flaw in this `moving-potential' model of thermodynamics: not only does it not conserve momentum, \emph{it does not conserve energy either}. So the assumption that the work done on the system equals the change in energy of the system is, in general, false; and so the basic assumptions of thermodynamics do not get off the ground.

This can be seen both formally and physically. On the formal side: except in the specific case where velocity is zero, the Hamiltonian of the moving-potential system is not time-translation invariant, and so by Noether's theorem that Hamiltonian is not a constant of motion. On the physical side: suppose that we do work the system, on a timescale fast compared with its equilibration timescale, to increase its energy by $\delta U$ and its momentum by $\delta \vctr{p}$. In general this changes the system's velocity, and so puts it out of equilibrium: its velocity no longer matches that of the box walls, and so it will collide inelastically with them. More specifically, if we transform to the frame of the box (in which momentum is not conserved but energy is) the work done on the system is $\gamma (\delta U - \vctr{v}\cdot \delta \vctr{p})$, so that the system will equilibrate with 4-momentum $(\gamma \delta U - \vctr{v}\cdot \delta \vctr{p},0)$. Transforming back to the moving frame, the energy is now $\gamma^2(\delta U - \vctr{v}\cdot \delta \vctr{p})$, so that the box has done additional work
\be
W_{box} = (\gamma^2 - 1) \delta U - \gamma^2 \vctr{v}\cdot \delta \vctr{p}= \gamma^2 (v^2 \delta U - \vctr{v}\cdot \delta \vctr{p}).
\ee
Only in the special case where $\delta \vctr{p}$=$\vctr{v}\delta U$ --- in other words, when the work done is done in such a way as to leave the velocity unchanged --- does this additional work vanish. (More generally, it vanishes if we also adjust the control parameter as we do work, so that $\delta \vctr{v} = \delta (\vctr{P}/U).$ But there is no particular reason why an agent interacting with the system need be so constrained. And so the moving-potential model after all fails to describe a thermodynamic system.

It is again helpful to consider a non-relativistic analogy. Consider a fluid confined to a \emph{stationary} box (so, normally, modeled by a Hamiltonian dependent on a parameter representing system volume) and suppose that we want to treat pressure, not volume, as the control parameter. Work done on the system will in general change its pressure, and so if we want to keep the control parameter fixed at some value $P_0$, any such work done must be accompanied by an adjustment of the volume; that is, the Hamiltonian must be time-dependent during the period of work and subsequent equilibration. (In physical terms: the box must be expanded to keep the pressure constant, and doing so causes the system to do work on the walls of the box, partially counteracting the original work done.) So while formally nothing stops us (at least locally) solving for volume in terms of pressure and energy, using this to write the energy as a function of entropy and pressure, and differentiating to get
\be
\mathrm{d}U = \left(\pbp{U}{S}\right)_P\mathrm{d}S + \left(\pbp{U}{P}\right)_S\mathrm{d}P\equiv T'\mathrm{d}S + V'\mathrm{d}P
\ee
the resultant expression cannot be interpreted as a version of the First Law and set equal to external work done.

That is not to say that thermodynamics has no use for systems of fixed pressure; to the contrary, they are ubiquitous in chemical physics. But different machinery is used to treat them: we suppose that our system is in mechanical (but not thermal) equilibrium with a \emph{pressure bath}, an extremely large reservoir system at some pressure $P$ (for instance, perhaps the two systems are separated by a thermally insulating membrane which is free to expand or shrink. If some finite  work $W$ is done on the original system, it will expand by some amount $\Delta V$ in order to keep its pressure constant at $P$, doing work $P \Delta V$ on the reservoir, so that its total change in energy is
\be
\Delta U = W - P \Delta V.
\ee
Rearranging, we have 
\be
W= \Delta U + P \Delta V = \Delta (U+PV).
\ee
If we define the \emph{enthalpy} $E=U+PV$, we have an alternative form of the First Law applicable for systems at constant pressure: work done equals change in enthalpy, or in differential form,
\be
\stkd W = T\mathrm{d}S + V \mathrm{d}P.
\ee
Of course, if we have \emph{two} pressure reservoirs, at \emph{different} pressures (and assuming that the total volume of the two reservoirs is fixed), there is an additional source of work available: all we need to do is transfer volume $\Delta V$ from the lower-pressure to the higher-pressure reservoir, extracting additional work 
\be 
W=(P_{\mbox{high}}- P_{\mbox{low}})\Delta V.
\ee

Returning to relativity, we can by analogy model a system constrained to move at velocity $\vctr{v}$, through dynamical interaction with some `velocity bath': a much larger reservoir system moving at that velocity. Suppose we do work $W$ on the system. The reservoir must transfer some quantity of momentum $\Delta \vctr{p}$ to the system in order to keep its velocity fixed, doing additional work $\vctr{v}\cdot \Delta \vctr{p}$, so that we have
\be
\Delta U = W + \vctr{v}\cdot \Delta \vctr{p}.
\ee
If we define the \emph{velocity enthalpy} $E_V$ as $E_V=W - \vctr{v}\cdot \vctr{P}$, we have a version of the second law appropriate for energy exchange with a system in contact with a velocity reservoir: work done equals change in velocity enthalpy, or in differential form,
\be\label{enthalpy-velocity}
\stkd W = \mathrm{d}E_V =T \mathrm{d}S - P\mathrm{d}V_r- \vctr{v}\cdot \mathrm{d}\vctr{p}= \frac{T_r}{\gamma}\mathrm{d}S- \frac{P_r}{\gamma}\mathrm{d}V_r - \vctr{p}\cdot \mathrm{d}\vctr{v}
\ee
(notice that the temperature appearing here is still that defined in the momentum-transfer model).

In fact, we can give an explicit expression for velocity enthalpy, which establishes a simple relation with (\ref{alt-first-law}). Since $\vctr{p}=\vctr{v}U$, we have
\be
E_V = (1-v^2)U = U/\gamma^2.
\ee
Differentiating, we get
\be
\mathrm{d}E_V = \frac{1}{\gamma^2}\mathrm{d}U - 2U \vctr{v}\cdot \mathrm{d}\vctr{v}.
\ee
If we substitute in (\ref{alt-first-law}) for $\mathrm{d}U$, we get
\begin{eqnarray}
\mathrm{d}E_V &=& \frac{1}{\gamma^2}(\gamma T_r \mathrm{d}S - \gamma P_r \mathrm{d}V_r + \gamma^2 \vctr{p}\cdot \mathrm{d}\vctr{v}) - 2 \vctr{p}\cdot \mathrm{d}\vctr{v}\nonumber \\
&=& \frac{T_r}{\gamma}\mathrm{d}S- \frac{P_r}{\gamma}\mathrm{d}V_r - \vctr{p}\cdot \mathrm{d}\vctr{v},
\end{eqnarray}
exactly as we obtained in (\ref{enthalpy-velocity}). From this perspective, the factor of $\gamma^2$ difference between the thermodynamic potentials found in the moving-potential model and in our original model can be tracked to the ratio of $\gamma^2$ between energy and velocity enthalpy.

\section{Temperature or rest temperature?}\label{rest-temperature}

There is a more interesting --- though ultimately still semantic --- proposal for how we might redefine the word `temperature' in the relativistic setting: namely, we might decide to take `temparature' as synonymous with `rest temperature', the thermodynamic temperature of a system in its own rest frame. After all, in general for a thermodynamic system there is no particular constraint on the equation of state --- given, say, conserved quantity $N$, pretty much any\footnote{In many cases, there is one physically-motivated constraint: extensivity, the requirement that entropy is first-order homogeneous in its arguments. But even this is not an in-principle system: self-gravitating systems violate it, as do systems small enough for edge effects to be significant.} well-behaved function $S=S(U,N)$ defines a thermodynamic system, and the particular function depends on the detailed physics of the system. But given a moving system characterized by energy $U$ and momentum $\vctr{p}$ (and perhaps by rest volume $V_r$), if the system is relativistically covariant then the equation of state has to have the form
\be
S=S(M(U,\vctr{p}),V_r)\equiv S(\sqrt{U^2-\vctr{p}\cdot\vctr{p}},V_r)
\ee
so that all the interesting physics is contained in the function $S(M,V_r)$, with dependence on $U$ and $\vctr{p}$ separately following just from Lorentz covariance.  From that point of view, the physically interesting quantity is $T_r \equiv \mathrm{d}M/\mathrm{d}S$, and one might decide to use `temperature' to refer to that quantity. Indeed, that is the normal convention in astrophysics and cosmology: `the temperature', used of an astrophysical body, in most circumstances refers to its temperature in a co-moving frame, \iec its rest temperature. (On occasion, `temperature' instead means `radiation temperature', as defined in section~\ref{covariant-thermodynamics}.)

One might indeed decide on that redefinition. But it is important to note that it is a \emph{redefinition}, a choice of how to redefine our terminology. The basic concepts of thermodynamics remain directly and unproblematically applicable to relativistically moving systems, and for those systems, thermodynamic temperature remains well defined and not in general equal to rest temperature.

By analogy, consider the notion of rest \emph{mass} in relativistic mechanics. In (one form of) \emph{Newtonian} mechanics, the Second Law is that force $\vctr{F}$ equals rate of change of momentum $\vctr{p}$ with time and that momentum is mass times velocity, \iec, mass is the ratio of velocity to momentum. Thus expressed, there is nothing relativistic about Newtonian mechanics: the Second Law does not itself make any presumptions about the symmetry structure of spacetime (at least with respect to velocity boosts) and the difference between nonrelativistic and relativistic mechanics is that in relativistic mechanics a body placed in motion at velocity $\vctr{v}$ has its mass increased by a factor $\gamma$, so that the Second Law for a relativistic system can be written as
\be
\vctr{F}= \dbd{\vctr{p}}{t}=\dbd{(m_r\gamma\vctr{v})}{t}
\ee
(where $m_r$ is the rest mass of the body, \iec the mass as measured in a frame comoving with it), whereas for a nonrelativistic system it is instead
\be
\vctr{F}= \dbd{\vctr{p}}{t}=\dbd{(m_r\vctr{v})}{t}= m_r\dbd{\vctr{v}}{t}
\ee
 (See \cite[ch.3]{brownrelativity} for more on this point.) But precisely because mass --- and indeed force, and rate of change of momentum with coordinate time --- are not relativistically covariant, it is natural and helpful to define a \emph{four-force} $\tilde{F}=(\gamma \vctr{F}\cdot \vctr{v},\gamma\vctr{F})$ and a four-momentum $\tilde{p}=m_r(\gamma,\gamma \vctr{v})$, and reexpress the Second Law as the law that four-force equals rate of change of four-momentum with proper time and that four-momentum equals rest mass times four-velocity,
 \be
 \tilde{F}=\dbd{\tilde{p}}{\tau}=m_r \frac{\mathrm{d}^2x}{\mathrm{d}\tau^2}.
 \ee
 The physically interesting property of a given body is its rest mass, from which its inertial mass is simply determined, and the modern norm in relativity is to elide the `mass' in `rest mass', so that `mass' no longer refers to the (non-invariant) ratio of three-momentum to three-velocity, but to the (invariant) ratio of four-momentum to four-velocity. But it is important to remember that this is a change in definitions. There is nothing wrong with the original notion of inertial mass; it is just that it is often more convenient to work in a  more manifestly covariant way.

Similarly with temperature and rest temperature. The original notion of temperature --- rate of change of energy with entropy at constant momentum  --- remains well defined and is $1/\gamma$ times the rest temperature. Whether we choose to insist on the `rest' part of `rest temperature' or instead elide it is a harmless matter of definitions, to be decided on grounds of convenience; our choice has no implications for the validity of standard thermodynamics any more than eliding the `rest' of `rest mass' has implications for the validity of Newton's Second Law.

\section{The view from statistical mechanics}\label{stat-mech}

So far I have worked entirely in the framework of phenomenological thermodynamics, without consideration of microphysical foundations: this framework (\emph{pace} \cite{earman-relativistic-thermodynamics}) is rich enough to incorporate relativistic thermodynamics without any need to consider the microphysics explicitly. But phenomenological thermodynamics does have a microphysical foundation --- equilibrium statistical mechanics --- and that framework, too, requires no extension to relativity, since it already incorporates the statistical mechanics of a moving system as one special case among many.

 Specifically (and, for simplicity, specializing to quantum mechanics; the classical version is given \emph{mutatis mutandis}): a system in statistical mechanics is characterized by its Hamiltonian $\op{H}$ and by any other conserved quantities $\op{N}_i$ (both $\op{H}$ and $\op{N_i}$ might be functions of some external parameter(s) but for simplicity I ignore that possibility in this section). At equilibrium, it is described (perhaps up to some coarse graining) by the \emph{canonical distribution}\footnote{Where other conserved quantities than energy are present, this is sometimes called the \emph{grand canonical distribution.} It is also possible to formulate statistical mechanics in terms of microcanonical distributions, but in general the canonical formulation is easier to work with.},
 \be
 \op{\rho}(T,\mu^i)=\frac{1}{Z(T,\mu^i)}\exp{-\frac{1}{T}\left( \op{H} - \sum_i\mu^i \op{N}_i\right)}.
 \ee
 Here, $T$ and the $\mu^i$ are to be thought of as implicit functions of the expectation values $U$ and $N_i$ of $\op{H}$ and $\op{N}_i$. The link to phenomenological thermodynamics is established when we identify the thermodynamic entropy with the von Neumann entropy and the thermodynamic values of energy and the other conserved quantities with their expectation values on the canonical distribution. (For an explicit presentation of thermodynamics and equilibrium statistical mechanics in this form, see \cite{wallaceirreversibility}; for some worries about the strategy of identifying thermodynamic values with expectation values, see \cite{friggwerndlcansomeoneplease}; for a response to these worries, see \cite{wallaceyourewelcome}.)
 
 To obtain the statistical mechanics of moving systems, then, all we need to do is to write down the canonical distribution for a system with three conserved momenta $\op{\vctr{p}}$ along with energy:
 \be
 \op{\rho}(T,\vctr{v})=\frac{1}{Z(T,\mu^i)}\exp{-\frac{1}{T}\left( \op{H} - \vctr{v}\cdot \op{\vctr{p}}\right)}.
 \ee
 (The identification of velocity with the potential conjugate to momentum follows either by pulling our previous result back through the derivation of thermodynamics from statistical mechanics, or from direct calculation.)
 
Since Lorentz transformations are unitarily implementable and trace is invariant under unitary transformations, we immediately deduce that entropy and the partition function $Z$ are scalars. If we reintroduce the inverse 4-temperature from section~\ref{covariant-thermodynamics} and write $\op{p}^\mu$ for the components of the 4-momentum operator, we can rewrite the canonical distribution as
\be
\op{\rho}(\tilde\beta)=\frac{1}{Z(T_r)}\exp{\tilde\beta_\mu \op{p}^\mu}.
\ee
As an application of these results, consider a moving box of radiation. Since photons are non-interacting, we can treat each mode of the radiation field as a separate thermodynamic system, all at the same 4-temperature. For a mode comprising photons with wavevector $\tilde k=(k,\vctr{e}k)$, the 3-momentum operator is given by $\op{\vctr{p}}=\vctr{e}\op{H}$. The canonical distribution for that mode is then
\be
\rho_{\tilde k}\propto \exp -\frac{k\gamma}{T_r}(1- \vctr{e}\cdot\vctr{v})
\ee
--- that is, it is equal to the at-rest canonical distribution for that mode at the appropriate radiation temperature, matching our previous results.

\section{What has gone before}\label{what-has-gone-before}

The thermodynamical model of a moving relativistic system I presented in section~\ref{thermodynamics-of-moving-bodies} --- called the ``Planck-Einstein formulation'' by Liu~\citeyear{liueinsteinthermodynamics,liurelativisticthermodynamics}, whose account of the history I follow here --- is almost as old as relativity itself: it was proposed originally by Planck~\citeyear{planck1906,planck1907}, developed by \citeN{einsteinrelativisticthermodynamics} and \citeN{vonlauebook}, and codified in Pauli's \citeyear{pauliencyclopedia} and Tolman's \citeyear{tolmanrelativitybook} textbooks. But the derivation of that model is quite different. Planck, Einstein \emph{et al} are concerned from the outset with the appropriate transformation between the standard form of thermodynamics for a system at rest and an appropriate form for a system in motion. They are led to include a term $\vctr{v}\cdot \mathrm{d}\vctr{p}$ in the First Law by arguing that when work $\stkd W$ is done on a system in its rest frame, then the description of that same process in a moving frame must include some change of momentum so as to keep the system's velocity constant. We are then led to the expression
\be
\stkd W = P\mathrm{d}V - \vctr{v}\cdot \mathrm{d}\vctr{p}
\ee
which, when inserted into the schematic form of the First Law ($\mathrm{d}U=\stkd W + \stkd Q$) yields (\ref{antithesis-firstlaw-2}), modulo some further considerations about the relativistic transformation properties of volume and pressure. The underlying idea here --- which runs consistently through the literature --- is that we are engaged in building a new theory, whose ingredients are the thermodynamics of systems at rest and the transformation laws of special relativity, and that our confidence in that new theory comes from a combination of the evidence for the old theory and the argument that there is only one natural way to combine them. For instance, \citeN[153-4]{tolmanrelativitybook} states that
\begin{quote}
[t]he justification for using [the usual mathematical form of the First and Second Laws] as giving the content of the first and second laws of thermodynamics when applied to systems in a state of uniform motion, will depend on the fact that the transformation equations for the quantities involved will be such as to make the validity of those equations, in a set of coordinates with respect to which a thermodynamic system is in motion, equivalent to their validity in proper coordinates with respect to which the system is at rest. \textbf{In these latter coordinates, however, these expressions are merely a statement of the classical first and second laws for which we assume that there is adequate empirical justification}. (Emphasis mine.)
\end{quote}
Similarly, when the consensus behind the Einstein-Planck formulation broke down in the 1950s (ironically due in part to Einstein's own criticisms; see \citeN{liueinsteinthermodynamics} for discussion), the framework remained the same: how should the notion of work be \emph{generalized} from the antecedently understood case of stationary systems to the novel case of relativistically moving systems? Einstein, and independently \citeN{ott}, argued that there was after all no need to add the momentum term to relativistic work, and ended up with an alternative formalism which Liu calls the Einstein-Ott formalism; essentially, it uses what I called the `constant-velocity temperature' as temperature, and my expression (\ref{alt-first-law}) as the First Law. And  \citeN{landsbergreview} argued that there was no really satisfactory relativistic \emph{generalization} of the First Law, so that thermodynamics by its nature would make sense only in the rest frame of the thermodynamic system.

The literature since then has been voluminous and tangled (see \citeN{liurelativisticthermodynamics} and references therein for routes into it); if it has established anything, it is that if the name of the game is to find a relativistic \emph{extension} or \emph{generalization} of the thermodynamics of static systems then there is no unique answer, only a variety of conflicting intuitions leading to conflicting proposals. There then seem to be two available attitudes. The first (which is the nearest the physics literature has found to a consensus) is \emph{coexistence}: there is no unique way to relativistic 
thermodynamics, but the different proposals are interdefinable and \emph{a fortiori} empirically equivalent. The second (ably advocated in recent work by \citeN{chuaTfallsapart}; see also \cite{earman-relativistic-thermodynamics}) is \emph{disintegration}: different features of thermodynamics which run together in the nonrelativistic regime come apart in relativity, so that there is no really satisfactory way to do relativistic thermodynamics. (``T falls apart'', as Chua memorably puts it.)

But throughout this debate it is assumed that standard thermodynamics is the thermodynamics of static systems, and there is no good reason to think that. (At least, not given a modern perspective on thermodynamics; it lies beyond the scope of this paper to consider how thermodynamics was understood contemporaneously with Einstein, Planck \emph{et al}.) Of course the historical applications of thermodynamics were to such systems, but modern thermodynamics is set up to consider conserved quantities \emph{in general}, and conserved momentum is no more a problem for its formalism  than conserved particle number or charge. Any system which is translation-covariant will conserve momentum, and so any such system will require momentum as well as energy on its list of conserved quantities, and will define temperature as rate of change of energy with entropy at constant momentum. We rarely include conserved momentum in the mainstream practice of thermodynamics, because in the vast majority of the systems to which thermodynamics (and statistical mechanics) is applied, translation symmetry is broken, either explicitly (by the walls of a fluid's container) or spontaneously (as occurs in solid matter). But a system whose dynamics are Poincar\'{e} (or indeed Galilei) covariant must be translation covariant, and so its thermodynamics must be formulated in terms of momentum as well as energy, and from there it is simply a matter of formal calculation to establish the transformation properties of the thermodynamic potentials.

Those calculations reproduce the Einstein-Planck formulation of thermodynamics, and so the Einstein-Planck formulation of thermodynamics is the correct literal statement about the thermodynamics of moving bodies. Since standard thermodynamics treats energy and momentum quite asymmetrically in a way which can hide relativistic covariance, and since the operational significance of energy transfer at constant momentum is limited, we might well choose to pay attention to other measures of the covariation of energy and momentum with entropy, such as constant-velocity temperature, radiation temperature, or rest temperature, and in some circumstances we might even decide to use the word `temperature' to describe those measures. But that choice conceals no residual conceptual puzzles; it is simply a shallow matter of semantics.

\section*{Acknowledgements}

I am indebted to Eugene Chua, whose thoughtful recent work on relativistic thermodynamics motivated me to consider the issue and with whom I had several illuminating conversations. Thanks also to John Norton and to an anonymous referee for helpful comments on a previous version of this paper.


\end{document}